\documentclass[12pt]{article}
\usepackage{amsmath,amssymb,amsthm,amsxtra,overpic,bbm,bm,epsfig,ulem,multirow}
\usepackage{color,cite}
\usepackage{epstopdf}
\usepackage{subfigure}
\usepackage{color}
\usepackage{graphicx}
\usepackage{dcolumn}
\usepackage{bm}
\usepackage[dvipdfm,pdfstartview=FitH]{hyperref}
\usepackage{graphicx,graphics}
\hypersetup{
    colorlinks=true,
    linkcolor=blue,
    filecolor=gray,
    urlcolor=blue,
    citecolor=blue,
}
\usepackage[mathlines]{lineno}
\usepackage{subfigure}
\usepackage{lineno}

\def\thefootnote{\fnsymbol{footnote}}

\usepackage{array}

\begin{document}
\vspace{0.2cm}
\begin{center}
{\Large\bf Non-negligible Oscillation Effects in the Crustal Geo-neutrino Calculations}
\end{center}

\vspace{0.2cm}

\begin{center}
{\bf Ran Han$^{a}$}\footnote{Email: hanran@ncepu.edu.cn},
{\bf Yu-Feng Li$^{b}$}\footnote{Email: liyufeng@ihep.ac.cn}
and {\bf Xin Mao$^{a,b}$}\footnote{Email: maoxin@ihep.ac.cn}\\
{$^a$ Science and Technology on Reliability and Environmental Engineering Laboratory, Beijing Institute of Spacecraft Environment Engineering, Beijing 100094, China}\\
{$^b$ Institute of High Energy Physics, Chinese Academy of Sciences, Beijing 100049, China}
\end{center}

\vspace{1.5cm}

\begin{abstract}
  {An accurate prediction of the geo-neutrino signal from the crust serves as a necessary prerequisite in the determination of the geo-neutrino flux from the mantle. In this work we report the non-negligible effect associated to the exact three-flavor antineutrino survival probability in the calculation of the crustal geo-neutrino signal, which was usually approximated as a constant average in previous studies. A geo-neutrino signal underestimation of about 1-2 TNU is observed as a result of the oscillatory behaviour within the local crustal region extending for about 300 km from the experimental site.
  We also estimated that the Mikheyev-Smirnov-Wolfenstein matter oscillation  is responsible for a $0.1\%$-$0.3\%$ increase of the local crustal signal, depending on the detector location. This work reminds that the exact oscillation possibility in matter should be considered for future prediction of the local crustal geo-neutrino signal.}\\
\end{abstract}

\begin{flushleft}
\end{flushleft}

\def\thefootnote{\arabic{footnote}}
\setcounter{footnote}{0}
\newpage

\section{\label{intro}Introduction}
The electron antineutrinos $\overline\nu_{e}$ from the beta decays of heat generating elements (HGEs) (i.e., $^{}$U, $^{}$Th, $^{}$K)
in the Earth interior are usually called
geo-neutrinos~\cite{Eder-NP78-657-1966,Marx-CJP-B19-1969,Avilez:1980wi,Krauss:1983zn,Kobayashi-Fukao-GRS-18-1991,Rothschild:1997dd,Raghavan:1997gw}.
Geo-neutrinos are considered to be {a} unique probe to reveal the fundamental geochemical and geophysical issues~\cite{Fiorentini:2002bp,Fiorentini:2007te},
and have attracted increasing attention since the {experimental measurements from} KamLAND~\cite{KamLAND-Nature436-499-2005,Gando:2013nba,KamLANDnew}
and Borexino~\cite{Bellini:2010hy,Bellini:2013nah,Agostini:2015cba,Agostini:2019}.
The existence of geo-neutrinos was first indicated in 2005 by KamLAND~\cite{KamLAND-Nature436-499-2005}, and then first evidenced in 2010 by
Borexino~\cite{Bellini:2010hy}. The latest observations of geo-neutrinos have been obtained in KamLAND at the significance of 7.9 $\sigma$~\cite{KamLANDnew} and
in {Borexino} at 5.9 $\sigma${~\cite{Agostini:2019}}, respectively.
{Besides these two experiments, next-generation experiments like SNO+~\cite{SNO+:2006,SNO+:2016}, JUNO~\cite{JUNO1,JUNO2}, Hanohano~\cite{hano} and Jinping~\cite{jinp,Wan:2016nhe,jinp1} are also going to join the family of geo-neutrino observation.}
{Although the unambiguous signals of geo-neutrinos have been discovered and a certain amount of data has been collected by the pioneer detectors, more information is still required to make the discrimination between the signals from $^{}$U and $^{}$Th, observe the geo-neutrinos from the mantle~\footnote{A preliminary indication of the non-zero mantle signal at the 99.0$\%$ confidence level has been reported by the Borexino collaboration~\cite{Agostini:2019}},}
and test different models of the radiogenic heat.

{Among all the problems of geo-scientific importance, a direct observation of geo-neutrinos from the mantle~\cite{Fiorentini:2012yk} is vital to the determination of the mantle's radiogenic power and the mode of mantle convection~\cite{Nunokawa:2003dd,Fiorentini:2005mr,Dye:2010vf,Sramek:2012nk}.
However, different from the crust, the mantle is almost unreachable and we have very limited knowledge about the abundance and distribution of radioactive elements in this region.
The contributions of crust and mantle cannot be separated by the current and next-generation geo-neutrino experiments,
since they are liquid scintillator detectors, and rely on the inverse beta decay (IBD) detection process~\cite{Vogel:1999zy,Strumia:2003zx}, which is insensitive to the direction of low energy antineutrinos~\cite{Fields:2004tf}.
As a result, we are left with an indirect substraction method of extracting the mantle component of geo-neutrino events by subtracting the crustal geo-neutrinos from total experimental data~\cite{JUNO1,Huang:2013ggg,Huang:2014dpa,Strati:2014kaa,Takeuchi:2019fft,Gao:2019pvi,Reguzzoni:2019tir,Wan:2016nhe,Fiorentini:2012yk}.}
To carry out a precision geo-neutrino measurement for the mantle, the accurate calculation of crustal geo-neutrinos will be very important.

{The uncertainties of the crustal geo-neutrino calculations come from many aspects, including the experimental location,
geo-neutrino energy spectra at production, the oscillation probability and the interaction cross section at detection, among which two of them are substantial.
The first one comes from the uncertainty for the geological information in the crust,
in particular, for the structure, density and abundance of $^{}$U and $^{}$Th in the near-field crust of the experimental
site~\cite{Huang:2014dpa,Takeuchi:2019fft,Gao:2019pvi}.
The geochemical uncertainties of U, Th abundances are typically much larger than the geophysical uncertainties of the density and thickness.}
The second one is the contribution from antineutrino oscillations from the geo-neutrino production to
detection processes. The possibility for geo-neutrinos to keep their flavors unchanged during propagation is described by the survival probability.
{Although some studies~\cite{Reguzzoni:2019tir,Strati:2017,Strati:2018} did take into account of the three-flavor survival probability when doing the signal prediction.
In most the previous calculations, since one needs to integral the whole crust of the Earth,
an average survival probability is more likely to be approximated~\cite{KamLAND-Nature436-499-2005,Bellini:2010hy,Sramek:2012nk,Dye:2012,Fiorentini:2012yk,Huang:2013ggg,Huang:2014dpa,Strati:2014kaa,Takeuchi:2019fft,Gao:2019pvi}
to account for the geo-neutrino disappearance during their propagation:}
\begin{eqnarray}\label{average_p}
\left\langle P_{e e}\right\rangle \simeq \cos ^{4} \theta_{13}\left(1-\frac{1}{2} \sin ^{2} 2 \theta_{12}\right)+\sin ^{4} \theta_{13}\,.
\end{eqnarray}
Along with the improvement for the experimental accuracy of the oscillation parameters and current and future geo-neutrino measurements,
this simple approximation cannot be taken for granted and one should carefully treat the oscillation effects when making the crustal geo-neutrino calculations.

First, from the point of view of the crustal geology, the Earth is unevenly distributed. The crustal depth can range from 10 km in the oceanic region to more than 60 km in the vicinity of the Tibet Plateau.
Meanwhile, $^{}$U and $^{}$Th abundance distributions are also not uniform at the regional level and vary significantly according to their forming history.
The {combination} of these uneven distributions with the energy and distance dependent antineutrino oscillation probability
makes the average constant not a good approximation in the future accurate calculations.
{Second, the trajectories of geo-neutrinos during their propagation are inside the Earth, the terrestrial matter effects should also be considered for future precision } measurements~\cite{Wan:2016nhe,Li:2016txk,Capozzi:2013psa,Capozzi:2015bpa}.

{In this work, we plan to quantify the variations on the geo-neutrino signal caused by these oscillation effects and, particularly, demonstrate the former is non-negligible.}
Without involving any local refined geological models~\cite{Huang:2014dpa,Takeuchi:2019fft,Gao:2019pvi,Strati:2017} for one particular experimental site, we are going to use the global model of {\verb"CRUST1.0"}~\cite{CRUST1.0} to define the structural layers and density distributions of the whole crust, where {the crust is divided into} $1^{\circ}\times 1^{\circ}$ cells with a distinction between the oceanic and continental crusts.
Within {\verb"CRUST1.0"}, {the different crustal layers are subdivided according to a depth-dependent density profile.}
For the geo-neutrino calculation, we assign the average $^{}$U and $^{}$Th abundances for each crustal type according to the global average studies in Refs.~\cite{Huang:2013ggg,Rudnick:2013,White:2014,Strati:2017}.
We show that an underestimate of around 1-2 TNU for the geo-neutrino signal is observed for the current running and future detectors if the average constant is approximated in the calculations, among which most of the deviations happen in the local crustal region of around 300 km.

This work is organized as follows. In Sec.~\ref{osci} we give the calculation of geo-neutrino oscillations including the terrestrial matter effects. Then we calculate the local crustal geo-neutrino signals with exact and average antineutrino oscillation probabilities and discuss their effects in future precision experiments in Sec.~\ref{cal}.
{Finally, we summarize the results of this work in Sec.~\ref{conclude}.}

\section{\label{osci}Calculation inputs}

In this section, we first review the fundamentals of geo-neutrino production and detection, which are one of the prerequisites to calculate the geo-neutrino signals,
and then discuss the antineutrino oscillation effects when geo-neutrinos propagate from the production to the detection points. Because all the processes of antineutrino production, propagation and detection take place inside the terrestrial matter, we shall calculate the antineutrino survival probability in matter by assuming the averaged matter potential.

\subsection{Geo-neutrino production and detection}

{Geo-neutrinos are produced in the decay chains of the main natural radionuclides, which have half-lives longer than or compatible to
the age of the Earth~\cite{Fiorentini:2005mr}:} 
\begin{eqnarray}
^{238}\mathrm{U}\;&\rightarrow&\;{^{206}\mathrm{Pb}} +8\;{^{4}\mathrm{He}}+6\;e^{-}+6\;\overline\nu_{e} + 51.7\;[\mathrm{MeV}]\nonumber\,,\\
^{232}\mathrm{Th}\;&\rightarrow&\;{^{208}\mathrm{Pb}} +6\;{^{4}\mathrm{He}}+4\;e^{-}+4\;\overline\nu_{e} + 42.7\;[\mathrm{MeV}]\nonumber\,,\\
^{40}\mathrm{K}\;&\rightarrow&\;{^{40}\mathrm{Ca}}+\;e^{-}+\;\overline\nu_{e}+1.311\;[\mathrm{MeV}]\,.
\end{eqnarray}
The endpoints, which define the maximal energies that geo-neutrinos can carry, are 3.27 MeV, 2.25 MeV, and 1.311 MeV for the $^{238}$U, $^{232}$Th decay series and $^{40}$K beta decay, respectively.
{From Eq.(2) one can see that, for each radioisotope, there is a strict connection between the number of emitted geo-neutrinos and the radiogenic heat production rate. Thus geo-neutrino measurements might provide a significant check on the Earth content of heat generating elements.}

\begin{figure}[tp]
  \centering \includegraphics[width=0.5\columnwidth]{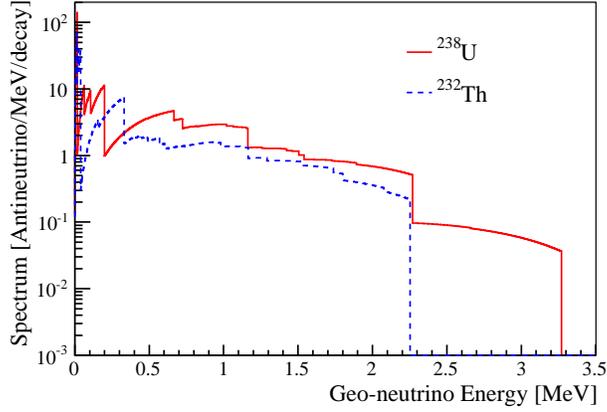}
  \caption{
            {\label{spectra}
            The geo-neutrino energy spectra of $\rm^{238}U$ and $\rm^{232}Th$. The spectra are normalized to the head elements of the decay chains and the integral of each spectrum corresponds to the number of geo-neutrinos produced in each decay chain.}
        }

\end{figure}

The energy spectrum for each beta transition with maximum electron energy $E_{\max}$ is well established~\cite{Enomoto:2005}:
\begin{eqnarray} \label{spec}
{N(E_e)} = \frac{{G_F^2{{\left| M \right|}^2}}}{{2{\pi ^3}}}F(Z,{E_e}){(E{}_{\max } - {E_e})^2}E_e\sqrt {E_e^2 - m_e^2}\,,
\end{eqnarray}
{where $G_{F}$ is the Fermi constant. $F(Z,{E_e})$ is the Fermi function accounting for the effect of the coulomb field of the nucleus and $Z$ is the nuclear charge of the daughter nucleus. The antineutrino spectrum is obtained from the beta spectrum by replacing ${E_{\overline \nu }} = {E_{\max }} - {E_e}$, according to the energy conservation in the beta decay process.}
The whole antineutrino spectra for $^{238}\mathrm{U}$ and $^{232}\mathrm{Th}$ can be obtained as shown in Fig.~\ref{spectra}, {by summing up the normalized energy spectra of every beta transition contained in the decay chains of the isotopes:
\begin{eqnarray} \label{spec1}
{f_X}({E_{\overline \nu }})=\sum\limits_{ij} R_{ij}\sum\limits_{k}I_{ij,k}\left(\frac{1}{K}N_{ij,k}(E_{\overline \nu})\right)\,,
\end{eqnarray}
where $X$ represents $^{238}\mathrm{U}$ or $^{232}\mathrm{Th}$, $R_{ij}$ is the probability of beta decay $i\to j$ in the chain, the value of which can be recursively obtained according to the branching ratios of the decay series. $I_{ij,k}$ is defined as the percentage intensity of the $k$th beta transition in the beta decay $i\to j$, particularly, $\sum\limits_{k}I_{ij,k}=1$. $K$ is the normalization constant that makes the integral of the spectrum $N_{ij,k}(E_{\overline \nu})$ equal to 1.}
The endpoint energies and the absolute intensity information are taken from the web site of the \href{https://www.nndc.bnl.gov/nudat2/}{National Nuclear Data Center} (NNDC). Compared to the previous widely used geo-neutrino spectra~\cite{Enomoto:2005}, our spectra are calculated using the latest nuclear database with more beta transition branches, i.e., 113 individual beta decays for $\rm^{238}U$ and 83 for $\rm^{232}Th$. 

{Geo-neutrinos are primarily detected by the IBD reaction, where an electron antineutrino interacts with a free proton in the organic liquid scintillator (LS) to produce a positron and a neutron:
\begin{eqnarray} \label{spec1}
\overline\nu_{e}\;+\;p \to e^{+}\;+\;n\;-\;1.806\;[\mathrm{MeV}]\,.
\end{eqnarray}
The positron quickly annihilates and releases two 0.511 MeV gamma rays which form a prompt signal. With a time interval of $\sim 200 \rm\mu s$, the neutron is captured by a proton and produce a 2.2 MeV gamma ray which forms a delayed signal.}
The prompt and delayed signals, which are correlated in position and time, making good distinguishing characteristics for electron antineutrinos. The threshold energy of the IBD reaction is approximately $\rm1.806$ MeV, which means there is no signal for the geo-neutrinos below this threshold and thus the antineutrinos produced by the $\rm^{40}K$ decay chains are invisible to the detector.

\subsection{Geo-neutrino oscillations}

The three-flavor survival probability for electron antineutrino $\overline \nu_e$ propagating in the vacuum can be described as~\cite{Li:2016txk}
\begin{eqnarray} \label{p}
{P_{ee}}= 1- P_{0} - P_{*}\,,
\end{eqnarray}
with
\begin{eqnarray} \label{p_delta}
P_{0}&=&{\sin ^2}2{\theta _{12}}{\cos ^4}{\theta _{13}}{\sin ^2}\Delta _{21}\nonumber\,,\\
P_{*}&=&\frac{1}{2}{\sin ^2}2{\theta _{13}}(1-{\cos }\Delta _{*}{\cos }\Delta _{21}+{\cos }2{\theta _{12}}{\sin }\Delta _{*}{\sin }\Delta _{21})\,,
\end{eqnarray}
and
\begin{eqnarray}
{\Delta _{ij}} &=& 1.27(\Delta m_{ij}^2L)/{E_{\overline \nu }}\nonumber\,,\\
  \Delta _{*}&=& \Delta _{31}+\Delta _{32}\,,
\end{eqnarray}
where $\theta _{12}$ and $\theta _{13}$ are the mixing angles, $\Delta m_{ij}^2$ is the square mass difference of antineutrinos in eV$^2$, $E_{\overline \nu}$ is the antineutrino energy in MeV, and $L$ is the propagation distance in meters. The values for the mixing angles and mass-squared differences have been measured by series of neutrino experiments.
In the calculation, we assume the mass ordering to be normal~($m_1<m_2<m_3$) and take the oscillation parameters provided by the Particle Data Group (PDG)~\cite{PDG:2018}. The four parameters required in Eq.~(\ref{average_p}) and Eq.~(\ref{p}) as well as their uncertainties are: ${\sin ^2}{\theta _{12}}=0.307\pm0.013$, ${\sin ^2}{\theta _{13}}=(2.12\pm0.08)\times10^{-2}$, $\Delta m_{21}^2=(7.53\pm0.18)\times10^{-5}\rm eV^2$, $\Delta m_{32}^2=(2.51\pm0.05)\times10^{-3}\rm eV^2$.
With these parameters, the average survival probability in Eq.~(\ref{average_p}) can be evaluated as
$
\left\langle {{P_{ee}}} \right\rangle=0.55_{-0.01}^{ + 0.01}
$.
By contrast, the exact survival probability in the framework of three generations of antineutrinos are calculated by Eq.~(\ref{p}).

The Mikheyev-Smirnov-Wolfenstein (MSW) effect highlights that antineutrino oscillation  can be modified in matter~\cite{Wolfenstein:1978}.
Taking this effect into consideration, {some modifications would be applied to get the effective oscillation parameters~\cite{Li:2016txk,Enomoto:2005}:}
\begin{eqnarray} \label{matter_eff}
{\sin ^2}2\tilde{\theta}_{12} &\simeq& {\sin^2}2{\theta}_{12}\left(1-2\frac{A_{\rm cc}}{\Delta_{21}}{\cos}2{\theta}_{12}\right)\,, \nonumber\\
\Delta \tilde{m}_{21}^2 &\simeq& \Delta {m}_{21}^2+A_{\rm cc}{\cos }2{\theta}_{12}\,,\nonumber \\
\tilde{\Delta}_{*}&\simeq& \Delta_{*}+A_{\rm cc}\,.
\end{eqnarray}
where
\begin{equation}\label{Acc}
 A_{\rm cc} = 2\sqrt{2}G_{F}N_{e}E_{\overline \nu}\,,
\end{equation}
{and $N_{e}$ is the electron number density related to the matter density of the propagation trajectory with an electron fraction of $\sim0.5$.}
For the matter density, we take the average value as described in {\verb"CRUST1.0"}.
Note that the correction for ${\sin }^2{\theta}_{13}$ is at the level of $\mathcal{O}(10^{-4})$ and can be safely neglected.
With the effective oscillation parameters, the survival probability in matter $\tilde{P}_{ee}$ can be calculated the same way as that in vacuum by Eq.~(\ref{p}).

\subsection{The global crustal model}

{\verb"CRUST1.0"}~\cite{CRUST1.0}, an updated global model of crustal structure, is developed from the previous widely used model {\verb"CRUST2.0"} and characterized by the refined $1^{\circ}\times 1^{\circ}$ grid. The crustal thickness data are obtained from the active seismic techniques which can be used to image the Earth's structure.
Around 40 crustal types are assigned in {\verb"CRUST1.0"} and the crustal properties for areas without local seismic or gravity constraint can be extrapolated from them.
{Each of the} $1^{\circ}\times 1^{\circ}$ cells consists of 8 layers: water, ice, upper sediments, middle sediments, lower sediments, upper crust, middle crust and lower crust. Four parameters including boundary depth, compressional velocity, shear velocity and density are given for these layers.
To calculate the geo-neutrino flux, we still need a full knowledge of the distribution of radiogenic elements U and Th.  We take the average abundances from Refs.~\cite{Huang:2013ggg,Rudnick:2013,White:2014} and assume them to be uniform in each layer, see Table~\ref{abundance}. Here, the upper five layers in CRUST1.0 are treated as one single sedimentary layer and the crust is further divided into continental crust and oceanic crust.
\renewcommand\arraystretch{1.5}
\begin{table}[htpb]
\setlength{\abovecaptionskip}{4pt}
\setlength{\tabcolsep}{9.5mm}
\setlength{\belowcaptionskip}{4pt}\centering
\caption{\label{abundance}%
  {Summary of U, Th abundances (in ppm) in geological layers, including the sedimentary layer (Sed), upper crust (UC), middle crust (MC) and lower crust (LC). The crust also distinguishes between continental crust (CC) and oceanic crust (OC).}
}
\begin{tabular}{cccccc}
\hline
\hline
Element&Sed&\multicolumn{3}{c}{CC}&OC  \\
\cline{3-5}
  &&UC&MC&LC&\\
  \hline
  U&${\rm{1}}{\rm{.73}}$&${\rm{2}}{\rm{.70}}$&${\rm{1}}{\rm{.30}}$&${\rm{0}}{\rm{.2}}$&${\rm{0}}{\rm{.07}}$ \\
  Th&${\rm{8}}{\rm{.10}}$&${\rm{10}}{\rm{.5}}$&${\rm{6}}{\rm{.5}}$&${\rm{1}}{\rm{.2}}$&${\rm{0}}{\rm{.21}}$ \\
  \hline
  \hline
\end{tabular}
\end{table}

\section{\label{cal}Crustal geo-neutrino signal}

{In this section we start from the discussion about the geo-neutrino signal prediction method. Then we calculate the local crustal signal at different experimental sites both for the case of average survival probability and for the case of exact survival probability.
The correction on geo-neutrino signal given by the MSW oscillation is also discussed.}

\subsection{Geo-neutrino prediction}

In general, the differential geo-neutrino signal from isotope $X$ with energy between ${E_{\overline \nu }}$ and ${E_{\overline \nu }}+{dE_{\overline \nu }}$ can be calculated from
\begin{eqnarray} \label{diff_signal}
{S_X}({E_{\overline \nu }}) = {N_p}t\varepsilon ({E_{\overline \nu }})\sigma ({E_{\overline \nu }})\Phi {}_X({E_{\overline \nu }})\,.
\end{eqnarray}
Here, ${N_p}$ is the number of free protons in the liquid scintillator. $t$ is the detecting time and $\varepsilon ({E_{\overline \nu }})$ is the efficiency for the detector; $\sigma ({E_{\overline \nu }})$ is the cross section for IBD reaction taken from Refs.~\cite{Dye:2012};
{$\Phi {}_X({E_{\overline \nu }})$ is the differential oscillated flux for geo-neutrinos produced at position $\vec r$ to get to the detector at position $\vec R$:}
\begin{eqnarray} \label{flux}
{\Phi _X}({E_{\overline \nu }})= \int_{V_{\oplus}}\limits {\frac{{\rho (\vec r)}}{{4\pi {{\left| {\vec R  - \vec r } \right|}^2}}}}   \frac{{{\alpha _X}(\vec r){C_X}}}{{{\tau _X}{m_X}}}  {f_X}({E_{\overline \nu }})
{P_{ee}}({E_{\overline \nu }},\left| {\vec R  - \vec r } \right|)d\vec r\,,
\end{eqnarray}
{where $V_{\oplus}$ is the physical volume over which the integration is taken;} ${\rho (\vec r)}$ is the rock density at $\vec r$; $\alpha_X(\vec r)$, $C_X$, ${\tau _X}$ and ${m_X}$ are the elemental mass abundance, isotopic concentration, life time and mass of the nucleus for isotope X;
{${f_X}({E_{\overline \nu }})$ is the geo-neutrino energy spectrum normalized to the number of antineutrinos $n_{X}$ emitted per decay chain;}
${P_{ee}}({E_{\overline \nu }},\left| {\vec R  - \vec r } \right|)$ is the survival probability for electron-type antineutrinos with energy $E_{\overline v }$. In the calculation, the isotopic concentration is taken from Refs.~\cite{Meija:2016}. The nuclear property parameters like life time and mass of nucleus are taken from NNDC.

Then, the total geo-neutrino signal can be obtained by integrating the differential signal over the energy.
By assuming the detector with $100\%$ efficiency and $10^{32}$ target protons to operate continuously for one year, {the geo-neutrino signal can be expressed in the Terrestrial Neutrino Unit (TNU).}

The geo-neutrino flux is calculated based on discrete volume cells and each cell is assigned with geophysical and geochemical {attributes.}
{The flux from each cell is obtained by integrating over the cell volume, therefore the total differential flux $\Phi {}_X({E_{\overline \nu }})$ can be calculated by summing up the individual differential fluxes produced by all cells, layer by layer.}

{The two cases of} using average and exact survival probabilities are different in the way the flux of a single cell is calculated.
On the one hand, since the average survival probability $\left\langle {{P_{ee}}} \right\rangle$ is approximated as a constant, the term related to antineutrino energy ${f_X}({E_{\overline \nu }})$ can be taken out of the integral in Eq.~(\ref{flux}). For the flux from cell $i$ , there is only a need to calculate the 3-D integral on spatial coordinates:
\begin{equation} \label{grid}
\Phi^i_X = \frac{{{\alpha^i_X}\rho _i{C_X}}}{{4\pi {\tau _X}{m_X}}}\left\langle {{P_{ee}}} \right\rangle  {n_X}\int\limits_{i} {\frac{1}{{{{\left| {\vec R  - \vec r } \right|}^2}}}} d\vec r\,,
\end{equation}
where $n_X$ is the number of antineutrinos emitted by one nucleus of isotope $X$:
\begin{equation} \label{inte}
{n_X} = \int {{f_X}({E_{\overline \nu }}} )d{E_{\overline \nu }}\,.
\end{equation}
The total signal energy spectrum is thus given by
\begin{equation} \label{aveflu}
{S_X^{\left\langle {{P_{ee}}} \right\rangle}}({E_{\overline \nu }}) = \frac{{{N_p}t }}{{{n_X}}}\varepsilon ({E_{\overline \nu }}) \sigma ({E_{\overline \nu }})f_X({E_{\overline \nu }})\sum\limits_i {\Phi ^i_X}\,,
\end{equation}
On the other hand, the exact survival probability $P_{ee}$ is a function of energy and position.
As a result, an energy spectrum consisting of a serials of 3-D spatial integrals at specific energy $E_{\overline \nu }$ is required to be calculated for the flux of cell $i$:
\begin{eqnarray} \label{grid_p}
 {\varphi _X^i}({E_{\overline \nu }})=\frac{{{\alpha^i_X}\rho_i{C_X}}}{{4\pi {\tau _X}{m_X}}}  \int\limits_{i}{f_X}({E_{\overline \nu }}) {P_{ee}}({E_{\overline \nu }},\left| {\vec R  - \vec r } \right|)
  {\frac{1}{{{{\left| {\vec R  - \vec r } \right|}^2}}}}
  d\vec r\,.
\end{eqnarray}
And the total signal energy spectrum is:
\begin{equation} \label{grid_p}
{S_X^{P_{ee}}}({E_{\overline \nu }}) = {{N_p}t\varepsilon({E_{\overline \nu }}) \sigma ({E_{\overline \nu }}) }\sum\limits_i {{\varphi _X^i}({E_{\overline \nu }})}\,.
\end{equation}

\subsection{Calculation results}

The previous approximation of average oscillation effects is significantly prompted by the motivation that the oscillation lengths for geo-neutrinos are much smaller respect to their propagation distances~\cite{Dye:2012,boexino_dis}.
{The oscillation lengths are generally at a level of several tens of kilometers, depending on the geo-neutrino energy.}
Besides, it has been shown that almost half of the total geo-neutrino signal is generated in a regional crust within about 500 km from the {detector~\cite{kam_dis,Strati:2014kaa,Huang:2014dpa,Strati:2017}.}
Since the crustal depth, density and U, Th abundances are unevenly distributed, the oscillation effect of geo-neutrinos from this area should be treated with caution.
{This study focuses on the $10^{\circ}\times10^{\circ}$ local crust around the detector.}
Note that the general linear dimension of $1^{\circ}$ in the model {\verb"CRUST1.0"} is close to 100 km.
{To study signal contributions from smaller regions,} we subdivide the $1^{\circ}\times1^{\circ}$ tile into one hundred $0.1^{\circ}\times0.1^{\circ}$ cells with the same properties.

\begin{figure}[]
\centering

\subfigure[KamLAND]{
\begin{minipage}[c]{0.45\textwidth}
\centering
\includegraphics[width=6.5cm]{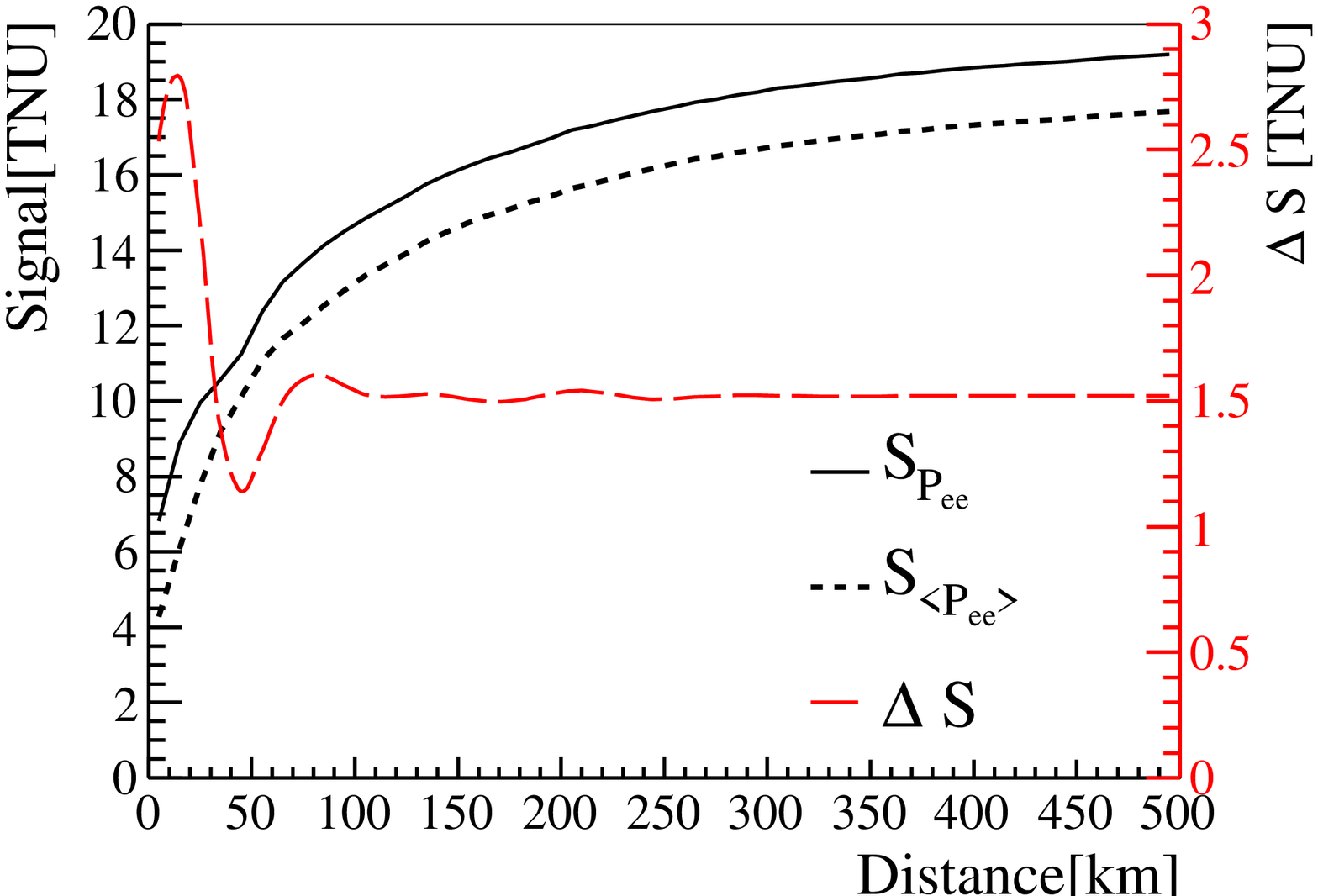}
\end{minipage}%
}
\subfigure[Borexino]{
\begin{minipage}[c]{0.45\textwidth}
\centering
\includegraphics[width=6.5cm]{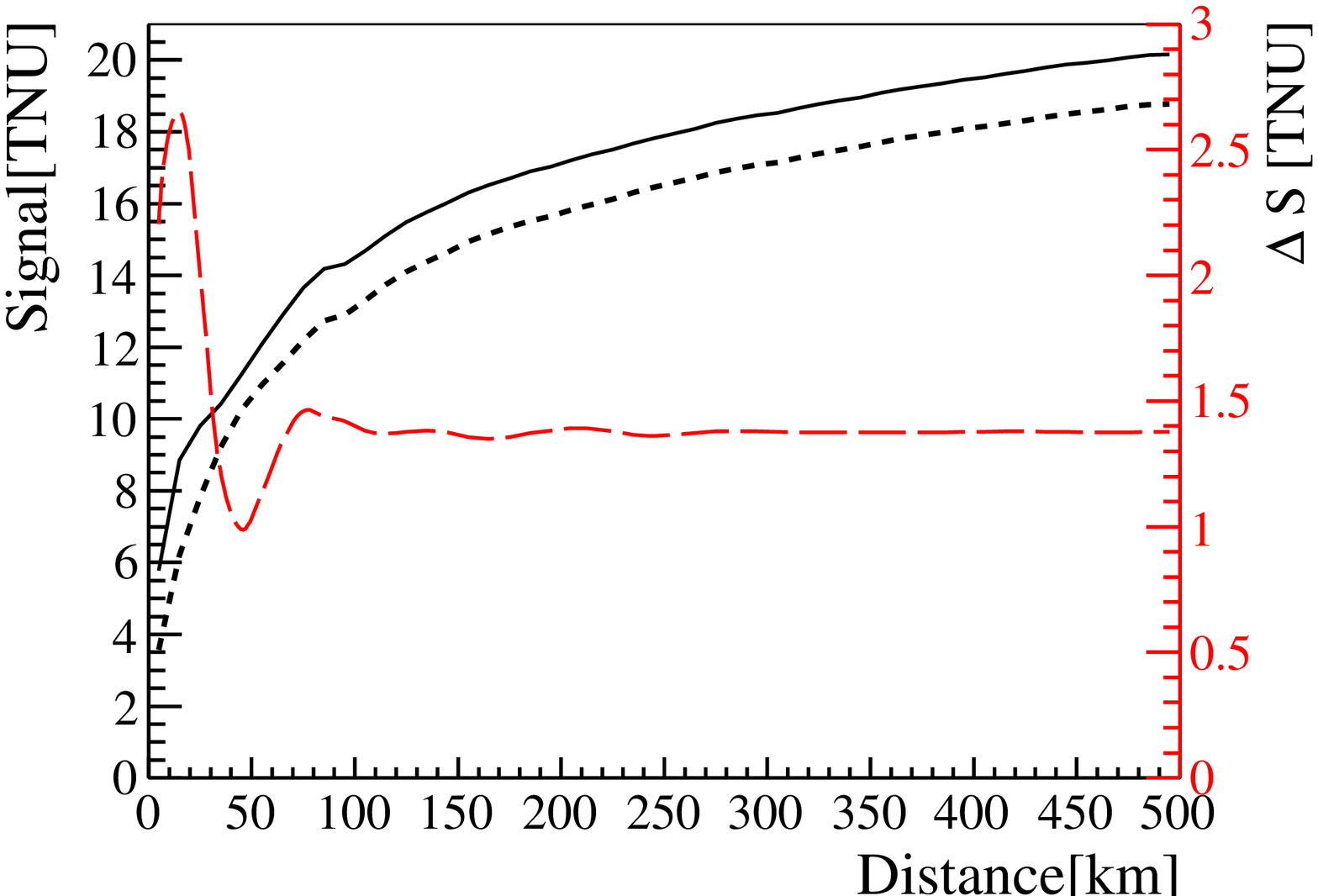}
\end{minipage}
}

\subfigure[SNO+]{
\begin{minipage}[c]{0.45\textwidth}
\centering
\includegraphics[width=6.5cm]{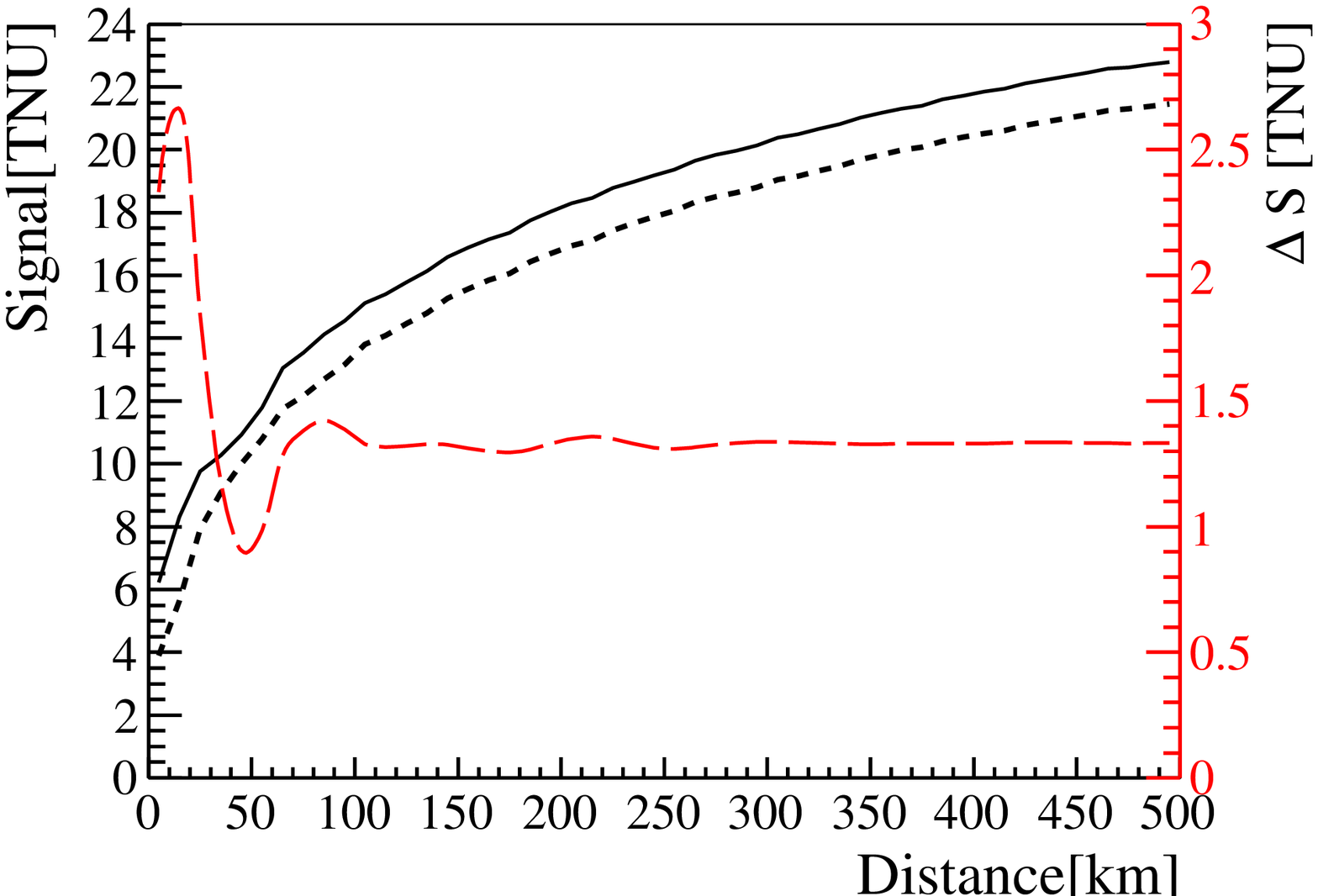}
\end{minipage}%
}
\subfigure[JUNO]{
\begin{minipage}[c]{0.45\textwidth}
\centering
\includegraphics[width=6.5cm]{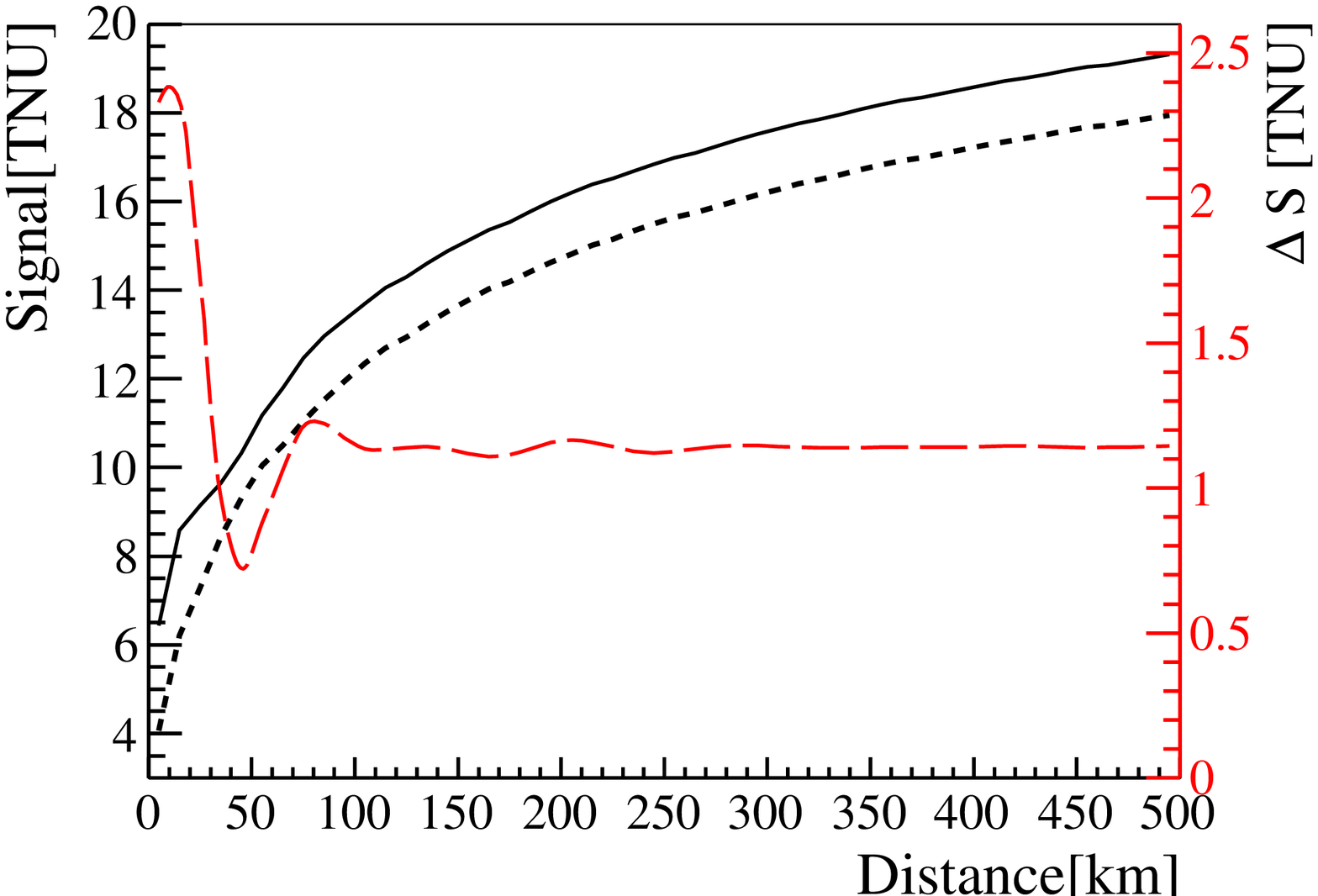}
\end{minipage}
}

\subfigure[Hanohano]{
\begin{minipage}[c]{0.45\textwidth}
\centering
\includegraphics[width=6.5cm]{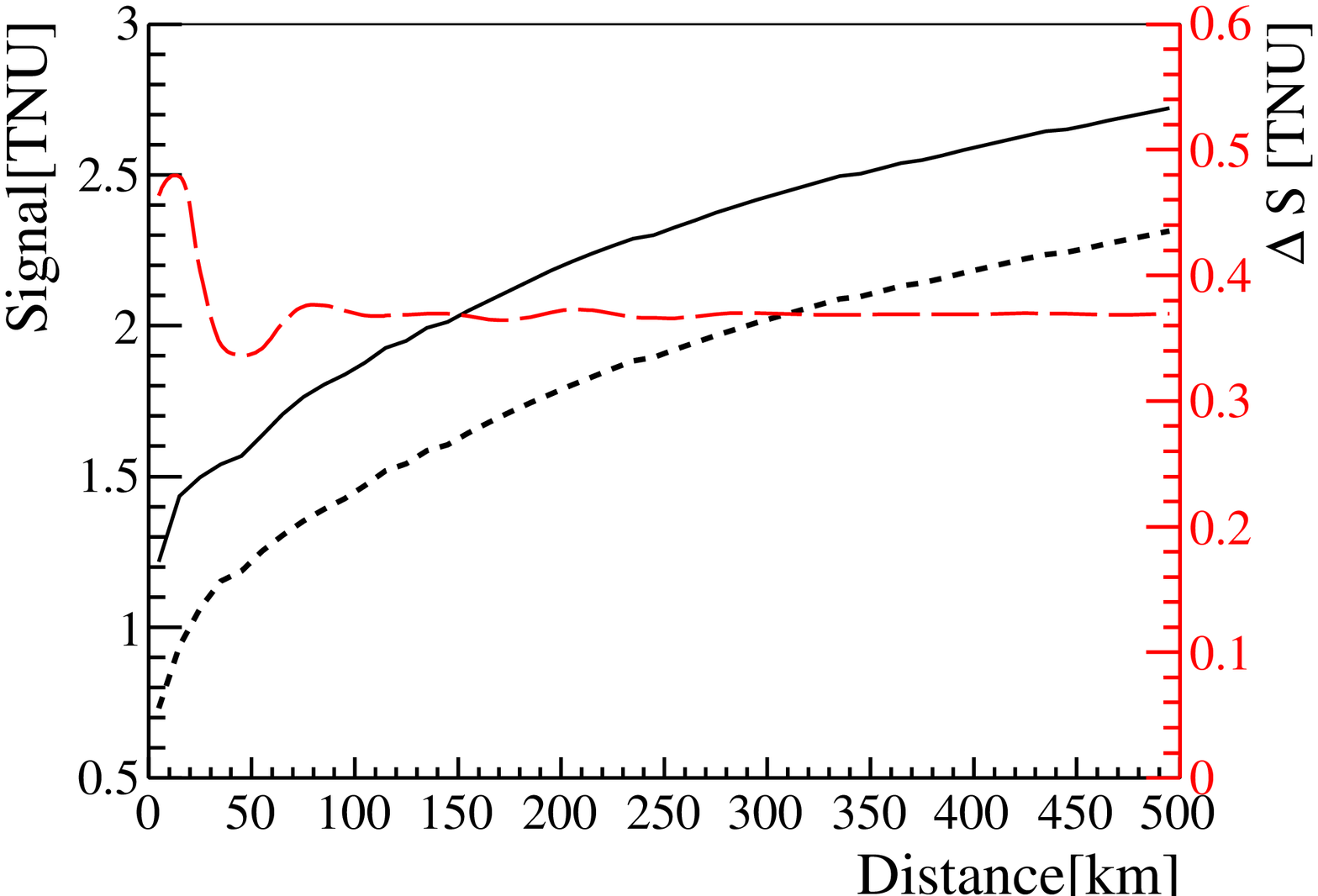}
\end{minipage}
}
\subfigure[Jinping]{
\begin{minipage}[c]{0.45\textwidth}
\centering
\includegraphics[width=6.5cm]{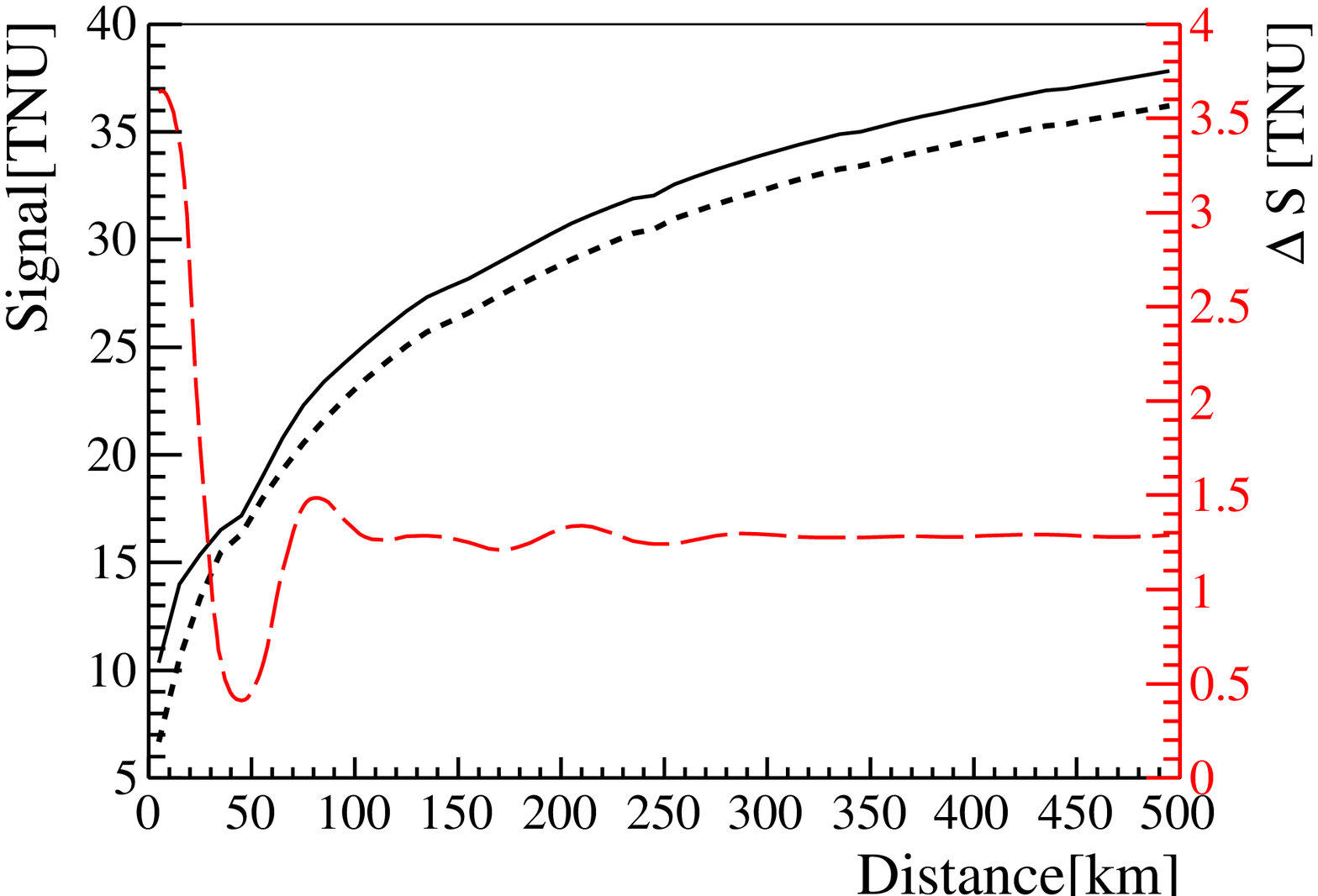}
\end{minipage}
}
\caption{
    \label{S_distance}
           { ¡°Cumulative signals calculated with ${P_{ee}}$ (black solid line, $S_{P_{ee}}$) and with $\left\langle {{P_{ee}}} \right\rangle$ (black dotted line, $S_{<P_{ee}>}$ ) together with their difference (red dashed line, $\Delta S=S_{P_{ee}}-S_{<P_{ee}>}$ ) as function of the distance from the detector for KamLAND (a), Borexino (b), SNO+ (c), JUNO (d), Hanohano (e), Jinping (f).} 
    }
\end{figure}

{The cumulative signals calculated with ${P_{ee}}$ and} $\left\langle {{P_{ee}}} \right\rangle$ and their difference as a function of distance from the detector are shown in Fig.~\ref{S_distance}. The cases of KamLAND, Borexino, SNO+, JUNO, Hanohano and Jinping are included. One can see that the using of the $\left\langle {{P_{ee}}} \right\rangle$ makes the signal predictions of different experiments reduced by a non-negligible amount.
As shown in Table~\ref{signal_pp}, the underestimated signals are between 1-2 TNU for most of the experiments except Hanohano with 0.41 TNU. Hanohano is designed to be a deep ocean detector, and the main geo-neutrino signal comes from the mantle due to the extremely thin crust at the experimental site. Therefore, the signal contribution of the crust and the resulting oscillation effect are very small for Hanohano.

One of the dominant sources of prediction uncertainty comes from the modeling of compositional variabilities (thickness, density and U, Th abundances).
{When calculating the total geo-neutrino signal based on general worldwide assumptions, the uncertainties related to the geophysical information of density and thickness are generally of the order of a few percent, while those associated to the U, Th abundances are typically more than $10\%$ (i.e. $17\%$ for KamLAND and $16\%$ for Borexino~\cite{kam_dis}).
As for the experimental measurements, the 1$\sigma$ relative uncertainties are 15.6$\%$ for KamLAND and 18$\%$ for Borexino respectively, according to their latest reports in Refs.~\cite{KamLANDnew,Agostini:2019}.}
The geology-associated uncertainties could be reduced by building a refined local crustal model around the {detector~\cite{Gao:2019pvi,Takeuchi:2019fft,Huang:2014dpa,Strati:2017}. }

\begin{table}[]
\setlength{\abovecaptionskip}{4pt}
\setlength{\tabcolsep}{2.0mm}
\setlength{\belowcaptionskip}{4pt}\centering
\caption{\label{signal_pp}%
  {Geo-neutrino signals calculated with ${P_{ee}}$ ($S_{P_{ee}}$) and $\left\langle {{P_{ee}}} \right\rangle$ ($S_{P_{<ee>}}$), the signal difference $\Delta S=S_{P_{ee}}-S_{<P_{ee}>}$ and the relative corrections given by MSW oscillation $100\times \frac{S_{\tilde P{_{ee}}}-S_{P_{ee}}}{S_{ P{_{ee}}}}$ for the local crustal regions of $10^{\circ}\times10^{\circ}$ centred at the locations of different existing and proposed experimental sites.}
} 
\begin{tabular}{p{2.5cm}<{\centering}p{4cm}<{\centering}p{2cm}<{\centering}p{1.8cm}<{\centering}p{1.8cm}<{\centering}p{3cm}<{\centering}}
\hline
\hline
  Experiment&Location&$S_{<P_{ee}>}$ [TNU]&$S_{P_{ee}}$ [TNU]&$\Delta S$ [TNU]&MSW Correction [$\%$]  \\
  \hline
  KamLAND&$36.43^{\circ}\rm N,137.31^{\circ}\rm E$~\cite{Enomoto:2005}&17.67&19.19&1.52&$0.20$\\
  Borexino&$42.45^{\circ}\rm N,13.57^{\circ}\rm E$~\cite{Bellini:2013nah}&18.78&20.16&1.38&$0.25$\\
  SNO+&$46.47^{\circ}\rm N,81.20^{\circ}\rm W$~\cite{SNO+:2006}&21.45&22.78&1.33&$0.27$\\
  JUNO&$22.12^{\circ}\rm N,112.52^{\circ}\rm E$~\cite{JUNO2}&17.95&19.31&1.36&$0.19$\\
  Hanohano&$20.00^{\circ}\rm N,156.00^{\circ}\rm W$~\cite{hano}&2.31&2.72&0.41&$0.14$\\
  Jinping&$28.20^{\circ}\rm N,101.70^{\circ}\rm E$~\cite{jinp1}&36.21&37.84&1.63&$0.24$\\
\hline
\hline
\end{tabular}
\end{table}

In this work, {we only consider the signal uncertainty coming from the uncertainties associated to the oscillation parameters entering the electron antineutrino survival probability.}
{Take JUNO as an example, and use the Gaussian Monte Carlo sampling of all the four oscillation parameters within their 1$\sigma$ ranges for 1000 times, the uncertainty of the predicted total local crustal signal, featured by the ratio of the 1$\sigma$ error over mean value, is estimated to be around $1.50\%$.
The contributions from every individual oscillation parameter to the signal uncertainty can also be obtained by sampling on the particular parameter when the other three parameters are fixed at mean values.
The resulting uncertainties are $1.47\%$ for ${\sin ^2}{\theta _{12}}$, $0.16\%$ for ${\sin ^2}{\theta _{13}}$, and $0.03\%$ for $\Delta m_{32}^2$ and $0.26\%$ for $\Delta m_{21}^2$, respectively.}
Obviously, the major contribution of uncertainty is from ${\sin ^2}{\theta _{12}}$ and a more precise measurement of this parameter is thus beneficial.

\begin{figure}[]
\centering \includegraphics[width=0.65\columnwidth]{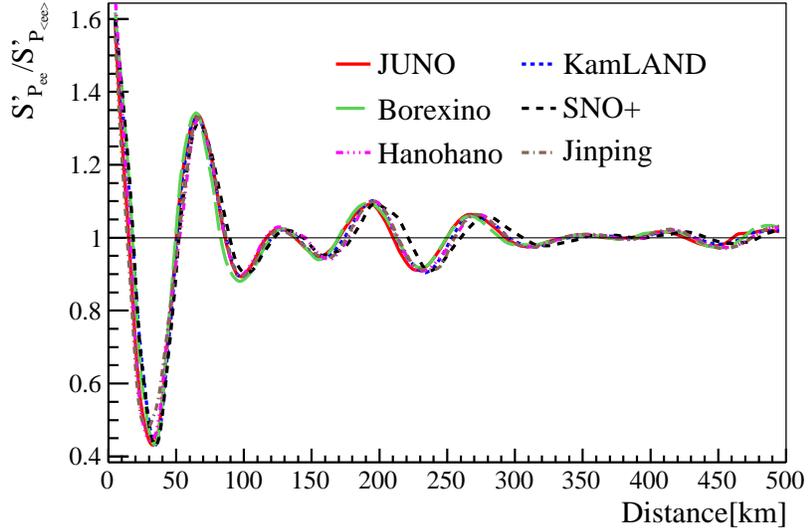}
\caption{
    \label{S_other}
        {Ratio $S'_{P_{ee}}/ S'_{<P_{ee}>}$ of the differential signals calculated respectively with ${P_{ee}}$ and $\left\langle {{P_{ee}}} \right\rangle$ as function of the distance from the detector for JUNO (solid red), Borexino (dashed green), Hanohano (dashed magenta), KamLAND (dotted blue), SNO+ (dashed black), Jinping (dashed grey).} 
    }
\end{figure}

{We define $S'$ as the differential geo-neutrino signal generated by U and Th distributed at a distance from the detector between $L$ and $L + dL$.}
Fig.~\ref{S_other} shows the ratio of differential signals calculated with ${P_{ee}}$ and $\left\langle {{P_{ee}}} \right\rangle$ as a function of distance {from} the detector. 
It should be noted that the difference only exist in the local area around 300 km, beyond this range, the difference will practically disappear.
{Thus we conclude that the average survival probability can be used for the far-field crust, i.e. at a distance greater than 300 km. Using the average survival probability in the near field would lead to an underestimation of the total signal between $2\%$ and $5\%$, depending on the detector location.}

\begin{figure}[]
\centering
\includegraphics[width=0.6\columnwidth]{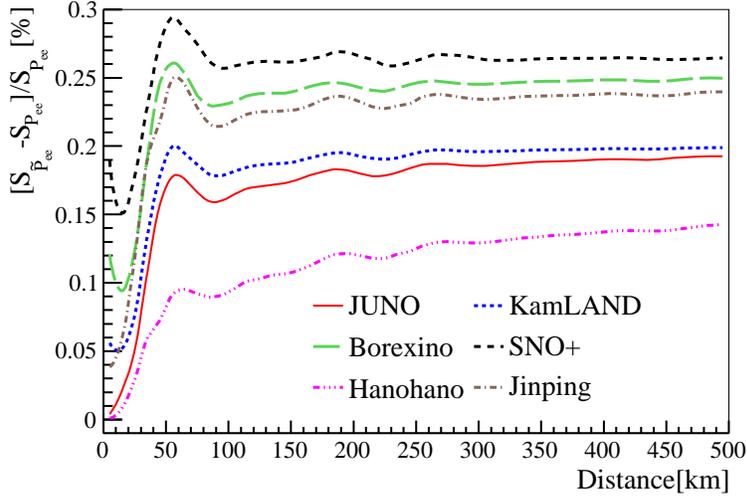}
\caption{
    \label{matter}
        {Relative difference between the cumulative signals ($S_{\tilde P{_{ee}}}$, $S_{P_{ee}}$) calculated respectively with the MSW oscillated survival probability (${\tilde P_{ee}}$) and the vacuum survival probability (${P_{ee}}$) as function of the distance from the detector for JUNO (solid red), Borexino (solid green), Hanohano (dashed magenta), KamLAND (dotted blue), SNO+ (dashed black), Jinping (dashed grey).}
    }
\end{figure}

Geo-neutrinos propagate through the Earth and will inevitably feel the matter potential of the electrons and nucleons. This is shown by the difference between the matter-corrected survival probability and the vacuum counterpart, of which the relative matter-induced correction can reach $4\%$ and is mainly attributed to $\Delta m_{21}^2$ and ${\sin ^2}{\theta _{12}}$~\cite{Li:2016txk}.
With ${\tilde P_{ee}}({E_{\overline \nu }},\left| {\vec R  - \vec r } \right|)$ to define the survival probability under the MSW oscillation effect, the relative difference in signal given by the MSW oscillation versus distance from the detector is shown in Fig.~\ref{matter}.
{We note that the effect of the MSW oscillation is mainly limited to a range of few hundred kilometres distance from the experimental site.}
{The MSW oscillation effect comes from a combined effect of the distance travelled by a geo-neutrino that is produced in a given geological reservoir with a specific matter density.}
The corrections are around $0.1\% $ to $0.3\%$ of the total local crustal signals, as shown in the rightmost column of Table~\ref{signal_pp}.
Clearly, the correction sizes for experimental sites dominated by continental crust (SNO+, Borexino and Jinping) are relatively larger. While for JUNO and KamLAND, which are on the margins of continents, the corrections are smaller. Hanohano has the least correction size because it {would} be located deep in the ocean.
{By using the averaged oscillation effects for the geo-neutrino signals \footnote{The geo-neutrino signals from the continental lithospheric mantle, depleted mantle and enriched mantle are inherited from Ref.~\cite{Strati:2014kaa}.} from the far field crust and mantle, the overall effect of the MSW oscillation on the total signal at JUNO is estimated to be $\sim0.3\%$.}

\section{\label{conclude}Conclusion}

In previous studies, the average oscillation effect was widely used for the prediction of total geo-neutrino flux. However, this will result in an inaccurate estimation for the near-field geo-neutrinos. For the mantle contribution indirectly obtained by predicting the crustal signal, this effect is particularly not negligible.

Based on the model {\verb"CRUST1.0"} and U, Th abundances provided by global geological studies, we calculate the local crustal signals using both the average and exact oscillation effects.
The results demonstrate that the average oscillation effect makes the predicted signal 1-2 TNU smaller than the exact one except for Hanohano with the signal underestimation of 0.41 TNU.
The signal deviations mainly exist in the near field of 300 km around the detector.
Therefore, it is better to carry out the accurate prediction with exact oscillation effect just within the region of local crust.
{Through Gaussian Monte Carlo sampling of the oscillation parameters, the uncertainty for the total local crustal signal is estimated to be $\sim1.5\%$, with the dominant contribution from ${\sin ^2}{\theta _{12}}$.}
For the MSW effect, the matter-induced corrections on the local crustal signal are estimated between $0.1\%$ to $0.3\%$ for different experimental sites,
{where the experimental sites in the continental crust have larger correction sizes than the counterparts located in oceanic crustal regions.}

{As we are going to enter the era of precision measurements of geo-neutrino signals,} in order to obtain an accurate and unbiased crust signal, we remind to use the exact survival probability within the local region of 300 km, while it is rather enough to approximate the calculation with the average survival probability in the far field area. This conclusion is even more relevant when the local refined crust models are employed.
We hope this study is useful for current and future geo-neutrino experiments.

\section*{Acknowledgements}

This paper is supported by National Key R\&D Program of China (2018YFA0404100), by the National Natural Science Foundation of China (Grants Nos.U1865206 and 11835013),
and by the Strategic Priority Research Program of the Chinese Academy of Sciences under Grant No. XDA10010100. Y.F. Li is also grateful for the support by the CAS Center for Excellence in Particle Physics (CCEPP).


\end{document}